\documentclass{amsart}
\usepackage{graphicx}
\usepackage{color}
\newtheorem{theorem}{Theorem}

  \newtheorem{remark}{Remark}

\UseRawInputEncoding
\usepackage[russian, english]{babel}

\def\dfrac#1#2{\displaystyle{#1\over #2}}

\def\zo{\omega}
\def\zt{\theta}

\def\bV{{\bf V}}

\def\zr{\rho}

\def\Div{\mbox{div}\,}

\def\bB{{\bf B}}

\def\bE{{\bf E}}

\begin{document}

\markboth{Rozanova, Chizhonkov}{On the influence of an external magnetic field on cold plasma}

%
%

\title{
The influence of an external magnetic field on cold plasma oscillations}

\author{Olga S. Rozanova*}

\address{ Mathematics and Mechanics Department, Lomonosov Moscow State University, Leninskie Gory,
Moscow, 119991,
Russian Federation,
rozanova@mech.math.msu.su}

\author{Eugeniy V. Chizhonkov}

\address{ Mathematics and Mechanics Department, Lomonosov Moscow State University, Leninskie Gory,
Moscow, 119991,
Russian Federation,
chizhonk@hotmail.com}



\maketitle


\begin{abstract}
For a system of equations describing one-dimensional nonlinear oscillations in a magnetoactive plasma, we study the effect of a constant magnetic field on the breaking of oscillations. For the nonrelativistic case, a criterion for the formation of a finite-dimensional singularity is obtained in terms of the initial data. It is shown that the enhancement of the magnetic field basically leads to an expansion of the class of initial data providing the global smoothness of the solution.
The nature of the singularities of the solutions is illustrated by numerical examples.
\end{abstract}


\section*{Introduction}	

\bigskip

Fully ionized plasma is a highly nonlinear medium in which even relatively small initial collective displacements of particles can lead to oscillations and large-amplitude waves. In the absence of dissipation, their evolution can lead to the appearance of a singularity of the electron density  \cite {Dav72}. This effect is commonly called {\it breaking} of oscillations and/or waves. As shown in  \cite {ZM},
singularity, i.e. the infinity of electron density in the Euler description of the motion of a medium is equivalent to the intersection of electron trajectories in its Lagrangian description. From a mathematical point of view, the breaking process means a blow up (the formation of infinite gradients of solution). %

The complexity and variety of problem formulations increases sharply if we consider the dynamics of a {\it magnetoactive} plasma, i.e. plasma placed in an external magnetic field. Even in the case of small disturbances, i.e. when considering only linear waves, a special classification is introduced in classical textbooks and monographs (see, for example,  \cite {ABR}, \cite {GR75}). In particular, waves propagating in the longitudinal and transverse directions with respect to the vector of a given magnetic field are distinguished. Already in a cold magnetoactive plasma, there are five oscillation branches propagating strictly along the external magnetic field, four of them describing transverse waves and one longitudinal waves \cite{Bernstein}, \cite {GR75}. It should be noted that, in the general case, linear longitudinal and transverse waves are not independent, as in a  isotropic (unmagnetized) plasma \cite {ABR}. It is for this reason that the modeling of nonlinear oscillations and waves in a magnetoactive plasma is of particular interest.  Numerical modeling  serves as the main research tool here, since the possibilities of the analytical and asymptotic approaches are severely limited due to the cumbersomeness and complexity of the equations describing the plasma dynamics. For example,  we can mention the monograph  \cite {CH18}, or more recent paper \cite {FCh21MM} devoted to the numerical simulation of oscillations in cold plasma, as well as wake waves excited by a short high-power laser pulse.

 Exact analytical results for the equations of cold plasma are very rare. However, they are the main theoretical value for the analysis of the acceleration of electrons in the wake wave of a powerful laser pulse  \cite {esarey09}. In  \cite {RChDAN20}, for the first time when using Eulerian variables, a necessary and sufficient condition for breaking plane one-dimensional oscillations in the nonrelativistic case was obtained; detailed proofs are given in  \cite {RChZAMP21}. Subsequently, the results of \cite {RChDAN20} were generalized to the case of accounting for electron-ion collisions  \cite {RChD20}.

In this paper, electrostatic 
oscillations of a magnetoactive plasma are considered with emphasis on the non-relativistic case. The methods described here make it possible to analyze in the relativistic case, but the results are rather cumbersome and we do not include them in this work in order not to overload the presentation.
It should be noted that earlier statements similar in meaning were previously considered only in the physical literature (see, for example, \cite {Dav72}, \cite {Kar16} and the references therein), where  methods and formulations of results differ significantly from ours.

The work is organized as follows. In Section \ref{Sec1}, a system of equations is derived that describes one-dimensional oscillations of an electron plasma in a constant magnetic field. In Section \ref{Sec2},  for the solution of the Cauchy problem for this system in the nonrelativistic approximation  we obtain  a blow up criterion in terms of the initial data. Hence follows the regularizing effect of a constant magnetic field in the general case. In Section \ref{Sec3}, we analyze subclasses of solutions, including traveling waves, for which the number of equations can be reduced.  
In Section \ref{Sec4}, for the purpose of numerical simulation of oscillations, we construct an algorithm is constructed in Lagrangian variables, which is closely related to the method of characteristics for the original differential setting. Numerical experiments clearly illustrate both the dynamics of smooth solutions and the formation of a gradient catastrophe.
Additional calculations confirm the stabilizing effect of the external magnetic field. In the conclusion, the results of the research are systematized.

\section{Statement of the problem on plasma oscillations}\label{Sec1}

We consider plasma as a non-relativistic electron liquid, neglecting recombination effects and ion motion. Then, in vector form, the system of hydrodynamic equations describing it, together with Maxwell's equations, will have the form:
\begin{equation}
\
\label{base1}
\begin{array}{l}

\dfrac{\partial n }{\partial t} + \Div(n \bV)=0\,,\quad
\dfrac{\partial \bV }{\partial t} + \left( \bV \cdot \nabla \right) \bV
=\frac {e}{m} \, \left( \bE + \dfrac{1}{c} \left[\bV \times  \bB\right]\right),\vspace{0.5em}\\
\dfrac1{c} \frac{\partial \bE }{\partial t} = - \dfrac{4 \pi}{c} e n \bV
 + {\rm rot}\, \bB\,,\quad
\dfrac1{c} \frac{\partial \bB }{\partial t}  =
 - {\rm rot}\, \bE\,, \quad \Div \bB=0\,,
\end{array}
\end{equation}
where
$ e, m $ --- charge and mass of the electron (here the electron charge has a negative sign: $ e <0 $),
$ c $ --- speed of light;
$ n,  \bV $ --- density and velocity of
electrons;
$ \bE, \bB $ --- vectors of electric and magnetic fields.
The system of equations ~ (\ref {base1}) is one of the simplest models of plasma, which is often called the equations of hydrodynamics of "cold" plasma;
it is well known and described in sufficient detail in textbooks and monographs (see, for example,  \cite {ABR}, \cite {GR75}).

Since the complete system is very difficult to study, physicists traditionally use various simplifications to study the effect of an external magnetic field. They come from the assumption of the smallness of various parameters, expansion in a series and retention of physically significant terms.  One of these models is a system that allows one to take into account the effects of an external magnetic field, which appeared, apparently, for the first time, in \cite{Dav68},
see also \cite{Dav72}, and has since emerged in various physical contexts, e.g. \cite{Kar16}, \cite{Maity12}, \cite{Maity13}, \cite{Maity13_PRL}. 
The solutions of this system are called in \cite{Dav68}
nonlinear zero-temperature Bernstein modes.

\begin{center}
\begin{figure}[htb]
\includegraphics[scale=0.5]{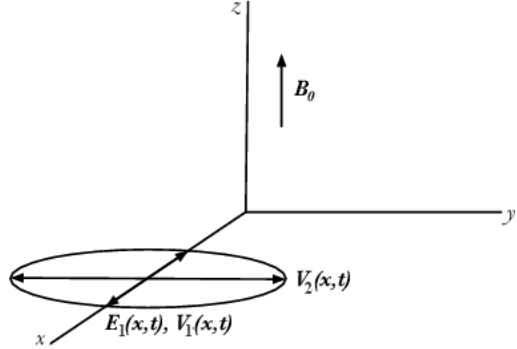}
\caption{The components of solution $V_1, V_2, E_1 $ and their dependence on coordinates}\label{Pic1}
\end{figure}
\end{center}

The system has the following form:
\begin{equation}
\begin{array}{c}
\dfrac{\partial n }{\partial t} +
\dfrac{\partial }{\partial x}
\left(n\, V_1 \right)
=0,\quad
\dfrac{\partial V_1 }{\partial t} + V_1 \dfrac{\partial v_1}{\partial x}= \frac{e}{m}\left[ E_1 + \dfrac1{c}\, V_2 B_0\right],
\vspace{1 ex}\\
\dfrac{\partial V_2 }{\partial t} + V_1 \dfrac{\partial V_2}{\partial x}= - \dfrac{e}{m}\,V_1  B_0,
\vspace{1 ex}\\
\dfrac{\partial E_1 }{\partial t} = - 4 \,\pi \,e \,n\, V_1.
\end{array}
\label{3gl2}
\end{equation}
Here
 $$
 \bV=(V_1(x,t), V_2(x,t), 0),\quad \bE(x,t) = (E_1(x,t),0,0),
$$
where $ (x, y, z) $ are the Cartesian coordinate.
It immediately implies that
 ${\rm rot}\, \bE = 0$ (the electric field is irrotational), in other words, the respective oscillations are electrostatic. The magnetic field does not depend on time and space in the model, it is directed along the axis $z$,   $\bB(x,t) = (0,0,B_0)$, $B_0 \equiv {\rm const}$, see Fig.\ref{Pic1}.
Despite its apparent simplicity,  system \eqref{3gl2} allows one to describe many interesting nonlinear effects.
We emphasize that it is not a direct consequence of \eqref{base1}, but uses additional assumptions.
System (\ref {3gl2}) is often called the equations of {\it upper hybrid} oscillations of cold magnetoactive plasma (see, for example,  \cite {Dav72}) in the sense that it describes the  dynamics of high Fourier harmonics and can be studied independently.  The hybridity is understood as a special type of perturbation of the plasma frequency by an external magnetic field.

For convenience, we introduce the dimensionless quantities
$$
\begin{array}{c}
\rho = k_p x, \quad \theta = \omega_p t, \quad
{\hat V_1} = \dfrac{V_1}{c}, \quad
{\hat V_2} = \dfrac{V_2}{c}, \quad
\vspace{1 ex}\\
{\hat E_1} = -\,\dfrac{e\,E_1}{m\,c\,\omega_p}, \quad
{\hat N} = \dfrac{n}{n_0}, \quad
{\hat B}_0 = -\,\dfrac{e\,B_0}{m\,c\,\omega_p},
\end{array}
$$
where $ \omega_p = \left (4 \pi e^ 2 n_0 / m \right)^{1/2} $ is the plasma frequency, $ n_0 $ is the value of the unperturbed electron
density, $ k_p = \omega_p / c $.
In the new variables, the system  (\ref {3gl2}) takes the form
\begin{equation}
\begin{array}{c}
\dfrac{\partial {\hat N} }{\partial \zt} +
\dfrac{\partial }{\partial \zr}
\left({\hat N}\, {\hat V}_1 \right)
=0,
\vspace{1 ex}\\
\dfrac{\partial {\hat V}_1 }{\partial \zt} + {\hat V}_1 \dfrac{\partial {\hat V}_1}{\partial \zr} =
 - {\hat E}_1 - {\hat V}_2 {\hat B}_0,
\quad
\dfrac{\partial {\hat V}_2 }{\partial \zt} + {\hat V}_1 \dfrac{\partial {\hat V}_2}{\partial \zr}=  {\hat V}_1 {\hat B}_0,
\vspace{1 ex}\\
\dfrac{\partial {\hat E}_1 }{\partial \zt} = {\hat N}\, {\hat V}_1.
\end{array}
\label{3gl3}
\end{equation}

From equations
$$
\dfrac{\partial {\hat N} }{\partial \zt} +
\dfrac{\partial }{\partial \zr}
\left({\hat N}\, {\hat V}_1 \right)
=0, \quad \dfrac{\partial {\hat E}_1 }{\partial \zt} = {\hat N}\, {\hat V}_1
$$
we have
$$
\dfrac{\partial }{\partial \theta}
\left[ {\hat N} +
\dfrac{\partial }{\partial \rho} {\hat E}_1 \right] = 0.
$$
This relationship is valid both in the absence of plasma oscillations and in their presence. Therefore, under the traditional assumption of a uniform background charge density of stationary ions, this implies a simpler expression for the electron density ${\hat N}(\rho,\theta)$:
\begin{equation}
 {\hat N}(\rho,\theta) = 1 -
\dfrac{\partial  {\hat E}_1(\rho,\theta) }{\partial \rho}.
\label{3gl4}
\end{equation}
Formula  (\ref {3gl4}) is a special case of the Gauss theorem  \cite {Dav72}, which in differential dimensional form is
$ \Div \bE =  4\,\pi\,e (n - n_0).$
Using  (\ref {3gl4}) in  (\ref {3gl3}), we arrive at the equations describing plane one-dimensional electrostatic 
plasma
oscillations. For convenience, we remove the "hat" from all dimensionless components of solution and obtain
\begin{equation}
\begin{array}{c}
\dfrac{\partial V_1 }{\partial \zt} + V_1 \dfrac{\partial V_1}{\partial \zr} =
 - E_1 - B_0\,V_2,
\quad
\dfrac{\partial V_2 }{\partial \zt} + V_1 \dfrac{\partial V_2}{\partial \zr}= B_0\, V_1,
\vspace{1 ex}\\
\dfrac{\partial E_1 }{\partial \zt} + V_1 \dfrac{\partial E_1}{\partial \zr} =  V_1.
\end{array}
\label{3gl5non}
\end{equation}



 We equip the equations (\ref {3gl5non}) with the initial conditions:
\begin{equation}\label{cd2}
     V_1(\rho,0) = V_1^0(\rho), \quad V_2(\rho,0) = V_2^0(\rho), \quad
     E(\rho,0) = E^0(\rho), \quad
 \rho \in {\mathbb R}.
\end{equation}

Completing the formulation of the problems, we recall the electron density formula useful for what follows:
\begin{equation}
 {N}(\rho,\theta) = 1 -
\dfrac{\partial  {E}(\rho,\theta) }{\partial \rho}.
\label{3gl4den}
\end{equation}

System  (\ref {3gl5non}) is of hyperbolic type. It is well known that for such systems, there exists, locally in time, a unique solution to the Cauchy problem of the same class as the initial data. 
 It is also known that for such systems the loss of smoothness by the solution occurs according to one of the following scenarios: either the solution components themselves go to infinity during a finite time, or they remain bounded, but their derivatives \cite {Daf16} turn to infinity. The latter possibility is realized, for example, for homogeneous conservation laws, which include the equations of gas dynamics, where the appearance of a singularity corresponds to the formation of a shock wave (a gradient catastrophe, \cite {RYa}).

\section{A criterion for the formation of a singularity of the Cauchy problem }\label{Sec2}

\begin{theorem} \label{T1}  For the existence of a $ C^1 $ smooth $ \frac {2 \pi} {\sqrt {1 + B_0^2}} $ - periodic solution \\ $ (V_1 (\theta, \rho), \, V_2 (\theta, \rho), \, E (\theta, \rho)) $ of the problem  (\ref {3gl5non}), (\ref {cd2}), $(V_1^0, V_2^0, E^0)\in C^2({\mathbb R}) $,
is necessary and sufficient that at any point $ \rho \in \mathbb R $ the condition
 \begin{equation} \label {crit2}
\Delta= \left( (V_1^0)' \right) ^ 2 + 2 \, (E^0)' +2 B_0\, (V_2^0)' - B_0^2  -1 <0.
\end {equation}
If the opposite inequality (\ref{crit2}) holds at least at one point $ \rho_0 $, then the derivatives of the solution turn to infinity in a finite time.
\end{theorem}

Proof.  Along the characteristics, the system \eqref {3gl5non} has the form
\begin{eqnarray}\label{char2}
     \dfrac{dV_1}{d\theta}=-E_1-B_0\,V_2,\quad \dfrac{dV_2}{d\theta}=B_0 \,V_1,\quad\dfrac{dE_1}{d\theta}=V_1,\quad \dfrac{d\rho}{d\theta}=V_1.
     \end{eqnarray}
This implies $V_1^2+V_2^2+E^2= V_1^2(0)+V_2^2(0)+E^2(0)=\rm const$ along each characteristic $\rho=\phi(\theta)$, starting from the point  $(\rho_0,0)$, and the solution remains bounded. Thus, the loss of smoothness by the solution can only be associated with the infinity of the derivatives.

Note that for a technical reason, to avoid further complication, we require a greater smoothness for the initial data. We denote $q_1=\partial_\rho V_1
$, $q_2=\partial_\rho V_2
$, $e=\partial_\rho E$, differentiate \eqref {3gl5non} with respect to $ \rho $ and get
\begin{equation}\label{char2d}
     \dfrac{dq_1}{d\theta}=-q_1^2-B_0 q_2 -e,\quad \dfrac{dq_2}{d\theta}=-q_1 q_2+B_0 q_1,\quad\dfrac{de}{d\theta}=(1-e)q_1,
     \end{equation}
which together with \eqref {char2} forms an extended system.

The positiveness of the density implies $ e <1 $, see \eqref {3gl4den}, therefore, below we consider only this part of the phase space. The case $ e = 1 $ corresponding to vacuum points must be considered separately.

System \eqref {char2d}  does not depend on \eqref {char2},
  it is integrable. Indeed, the last two equations imply
 \begin{equation}\label{C1}
   q_2=B_0+ C_1(e-1), \quad C_1=\frac{q_2(\rho_0,0)-B_0}{e(\rho_0,0)-1}.
     \end{equation}
Further, taking into account \eqref {C1}, from the first and third equations \eqref {char2d} we obtain
 \begin{eqnarray}\label{C21}
    C_2(e-1)^2=q_1^2+2(e-1)(B_0 C_1+1)+B_0^2+1, \\ C_2=   \frac{q^2_1(\rho_0,0)+2(e(\rho_0,0)-1)(B_0 C_1+1)+B_0^2+1}{(1-e(\rho_0,0))^2}.\nonumber
     \end{eqnarray}
  From \eqref {C21}, in particular, it follows that $ q_2 $ can go to infinity only together with $ e $.

On the plane $ (e, q_1) $, the algebraic equation \eqref {C21} defines a second-order curve that is bounded (an ellipse) if $ C_2 $, given as \eqref {C21}, is negative.
If we substitute here the value of $ C_1 $ from \eqref {C1}, then we get the condition
\eqref{crit2}.

Otherwise, the curve \eqref {C21}, the projection of the phase curve of the system \eqref {char2d} onto the plane $ (e, q_1) $, is a parabola or hyperbola, and the derivatives $ (e, q_1) $ go to infinity on it.

Let us show that this happens in a finite time.
Indeed, from \eqref {char2d}, \eqref {C21} it follows that $ e $ satisfies the equation
$$
     \dfrac{de}{d\theta}=\pm (e-1) \sqrt{C_2 (e-1)^2-2(e-1)(B_0 C_1+1)-B_0^2-1
     },
$$
    therefore
   \begin{equation}\label{razd2}
  \pm \int\limits_{e(\theta_*)}^{e(\theta^*)}  \dfrac{de}{(e-1) \sqrt{C_2 (e-1)^2-2(e-1)(B_0 C_1+1)-B_0^2-1}}= \theta^*-\theta_*,
     \end{equation}
   where $ (\theta _ *, e (\theta_ *)) $ is the starting point of integration, and $ (\theta^*, e (\theta^*)) $ is the final one.
    If the time for going to infinity is infinite, then  $ \theta^* = \infty $, $ e (\theta ^ *) = \pm \infty $, and the integral on the left side of \eqref {razd2} must diverge. However, it is easy to see that it is convergent. The resulting contradiction completes the proof for the case $ e <1 $.

If $ e = 1 $, then  system \eqref {char2d} is reduced to
\begin{equation*}
     \dfrac{dq_1}{d\theta}=-q_1^2-B_0 q_2 -1,\quad \dfrac{dq_2}{d\theta}=-q_1 q_2+B_0 q_1,
     \end{equation*}
its first integral under the assumption $ q_2 \ne B_0 $ has the form
\begin{equation*}
     q_1^2+2B_0 q_2 -B_0^2+1=C_3 (q_2-B_0)^2.
     \end{equation*}
This curve is defined on the phase plane $ (q_1, q_2) $ under the condition $ C_3 <0 $, which coincides with  \eqref {crit2} if we substitute in the last $ e = 1 $.

Finally, in the case $ e = 1 $, $ q_2 = B_0 $,   system \eqref {char2d}  reduces to one equation
\begin{equation*}
     \dfrac{dq_1}{d\theta}=-q_1^2-B_0^2 -1,
     \end{equation*}
the solution of which for any initial data turns to infinity in a finite time. It is easy to see that in this case   condition \eqref {crit2} cannot be met.

Formula \eqref {razd2} allows calculating the period $ T $ of revolution along the phase trajectory for $ C_1 <0 $. Namely,
$$
  \frac{T}{2}=\int\limits_{e_-}^{e_+}  \dfrac{de}{(e-1) \sqrt{C_2 (e-1)^2-2(e-1)(B_0 C_1+1)-B_0^2-1}},
$$
where $ e_ \pm $ is the smaller and larger root of the equation $C_2 (e-1)^2-2(e-1)(B_0 C_1+1)-B_0^2-1=0$.
  Standard calculations show that $ T = \frac{2 \pi}{\sqrt{1+B_0^2}} $ for all  $ C_1<0 $ and $C_2$.
 The same result follows from \eqref {char2}, since for smooth solutions
 $$\dfrac{d^2 V_1}{d \theta^2}+(1+ B_0^2)V_1=0. $$

Thus, the theorem is completely proved. $\Box$

\begin{remark} The criterion \eqref {crit2} coincides with  the criterion obtained in \cite {RChZAMP21} for the case $ B_0 = 0 $.

\end{remark}

\begin{remark} \label{R2}
From \eqref {crit2} it follows that if we fix {\it arbitrary} initial data (\ref {cd2}) and begin to increase the field $ B_0 $ in absolute value, then we get a globally smooth solution. That is, a constant magnetic field has a stabilizing effect (similar to the Coriolis force in geophysics \cite{Tadmor_Liu}, \cite{Rozanova_Usp}).
\end{remark}

\begin{remark} The condition \eqref{crit2} means that for any fixed $ B_0 $ an increase in
$\,  (V_2^0)' \,{\rm sign B_0}\,$ causes a blow up.
\end{remark}

\section{Subclasses of solutions and traveling waves in the nonrelativistic approximation}\label{Sec3}
\subsection{First integrals and subclasses of solutions}\label{Sec3.1}
As shown in \cite {RChZAMP21}, in the case $ B_0 = 0 $, $ V_2 = 0 $, there is a subclass of solutions distinguished by the condition
$E^2(\rho_0,0)+V_1^2(\rho_0,0)=K$, where $ K $ is a constant common to all points of $ \rho_0 \in \mathbb R $. In other words, the first integral of  equation \eqref {char2} was assumed to be equal to the same constant along all characteristics.

In our case, there are three equations, two functionally independent first integrals. They are, for example,
\begin{equation}\label{fe}
V_2-B_0 E_1 = K_1, \quad V_1^2+V_2^2+E_1^2=K_2.
\end{equation}
It is easy to see that for the case $ B_0 = 0 $, $ V_2 = 0 $, the integral $ K_2 $ corresponds to $ K $ from \cite {RChZAMP21}.

In addition, any differentiable function of $ K_1 $ and $ K_2 $ is also a first integral. This gives great opportunities for reducing the dimension of the system (reducing  to two or one equations).

A simple example is a system distinguished by the condition $ K_1 = 0 $, that is
\begin{equation*}
\dfrac{\partial V_1 }{\partial \zt} =
 - (1+B_0^2) E_1, \quad
\dfrac{\partial E_1 }{\partial \zt} =  V_1,
\label{odulin}
\end{equation*}
the solution with data \eqref{cd2} for which has the form
\begin{equation*}
\begin{array}{c}
E_1(\rho,\zt) = E_0(\rho)\, \cos(\zo\,\zt),\quad
V_1(\rho,\zt) = -\, E_0(\rho)\, \zo\, \sin(\zo\,\zt),
\label{odulin1}
\end{array}
\end{equation*}
with $ \zo = \sqrt{1+B_0^2}.$

\bigskip

\begin{remark}\label{R4}
It should be noted that the statement about the stabilizing effect of the magnetic field may not be valid for certain subclasses of solutions. For example, for the subclass considered above, distinguished by the condition $ K_1 = 0 $, from \eqref {crit2} we get
$$
\Delta=q_1^2+(2 e-1)(1+B_0^2),
$$
and at the points where $ e> \frac12 $, increasing $ B_0 $ will not make $ \Delta $ negative. However, this situation differs from the one described in Remark \ref{R2}, since the condition $ K_1 = 0 $ implies that  $B_0$  participates in the initial data and we cannot
increase $B_0$  fixing the initial data.
\end{remark}

\bigskip

\subsection{Traveling waves}

For the equations of cold plasma, solutions in the form of waves traveling with constant velocity (quasi-stationary solutions) are well known to physicists \cite {AL51, AP56}. In our work \cite {RChZAMP21} for the case $ B_0 = 0 $ we constructed  a subclass of solutions for which the first integral is constant.

  For the system  \eqref {3gl5non}, traveling waves can also be constructed. On such solutions both functionally independent first integrals are constant and for their initial data the condition $ \Delta <0 $ from \eqref {crit2} is necessarily satisfied.

Let us $V_1(\theta, \rho)={\mathcal V}_1(\xi)$, $V_2(\theta, \rho)={\mathcal V}_2(\xi)$,
$E(\theta, \rho)={\mathcal E}(\xi)$, $\xi=\rho-w \theta$, $w=\rm const$. From \eqref {3gl5non} we get the system
\begin{equation}\label{wave_sys}
(-w +{\mathcal V}_1){\mathcal V}_1\,'=-{\mathcal E}-B_0 {\mathcal V}_1,\quad
(-w +{\mathcal V}_1){\mathcal E}\,'={\mathcal V}_1, \quad   (-w +{\mathcal V}_1){\mathcal V}_2\,'=B_0{\mathcal V}_1.
\end{equation}
We choose the first integrals as 
\begin{equation}\label{fe_wave}
{\mathcal V}_2-B_0 {\mathcal E} = K_1, \quad 
 {\mathcal V}_1^2+{\mathcal V}_2^2+{\mathcal E}^2=K_2,
\end{equation}
whence follows
\begin{equation*}
{\mathcal V}_1^2+(1+B_0^2)\left({\mathcal E}+\frac{K_1 B_0}{1+B_0^2}\right)^2=K_2+\frac{K_1^2 B_0^2}{1+B_0^2}=K_3 \,(>0)
\end{equation*}
and
\begin{equation*}\label{wave_E}
{\mathcal E}=\pm\frac{1}{\sqrt{1+B_0^2}}\sqrt{{K_3}-{\mathcal V}_1^2}-\frac{K_1 B_0}{1+B_0^2}.
\end{equation*}
From \eqref {wave_sys}, \eqref {fe_wave} we get the equation for the traveling wave profile:
\begin{equation}\label{wave_P}
{\mathcal V}_1\,'   = \frac{\pm\sqrt{1+B_0^2}\sqrt{K_3-{\mathcal V}_1^2}}{-w +{\mathcal V}_1}.
\end{equation}
Since $${\mathcal V}_1^2\le K_3 \, (>0),$$
then under the condition
\begin{equation*}\label{w}
w^2> K_3
\end{equation*}
the denominator of \eqref {wave_P} preserves the sign and the function $ {\mathcal V} _1 $ at each half-period increases or decreases in $\xi $,
  changing between $ {\mathcal V}_\pm = \pm {K_3} $.
The period $ X $ can be calculated as
\begin{equation*}\label{wave_X}
X   =\frac{2}{\sqrt{1+B_0^2}} \,{\rm sign} \,w
 \int\limits_{-\sqrt{K_3}}^{\sqrt{K_3}} \frac{-w +\eta}{\sqrt{K_3-\eta^2}} \, d \eta=\frac{2\pi |w|}{\sqrt{1+B_0^2}}.
\end{equation*}
The profile $ {\mathcal V}_1 $ can be found implicitly as
\begin{equation*}\label{wave V}
 \left(\sqrt{1+B_0^2}\xi+c\right)^2= \left(w \arcsin \frac{{\mathcal V}_1}{\sqrt{K_3}} + \sqrt{K_3-{\mathcal V}_1^2}\right)^2,
  \end{equation*}
  $c= w \arcsin \frac{{\mathcal V}_1(0)}{\sqrt{K_2}} + \sqrt{K_3-({\mathcal V}_1(0))^2}$.

  \bigskip

\section {Numerical illustrations}\label{Sec4}

The approach to the analysis of characteristics used to obtain analytical results can be transformed into a high-precision approximate method for calculating nonrelativistic oscillations of a magnetoactive cold plasma. The specificity of the method lies in the identification of the function of displacement of particles (electrons) in Lagrangian variables with a function that characterizes the electric field in Euler variables. In fact, the method presented below is a method of characteristics; however, it is convenient to present its construction in terms of the Lagrangian
descriptions of the medium, i.e. using the concepts of particles and their trajectories.

Note that the simulated medium is represented by charged particles located in the background field formed by stationary ions. Therefore, for each particle there is an "equilibrium" position when the background field is canceled and which is convenient to take as the value of its Lagrangian coordinate. In this case, the  function of displacement relative to the equilibrium position is responsible for the formation of the electric field:
\begin{equation}\label{ptraek}
\rho(\rho^L,\theta) = \rho^L + R(\rho^L,\theta),
\end{equation}
where $ \rho ^ L $ is the "equilibrium" position of the particle when it does not contribute to the formation of the electric field, $ R (\rho ^ L, \theta) $ is its displacement function, which generates the electric field at the point of the trajectory $ \rho (\rho ^ L, \theta) $. According to ~ \cite {daw59} (see also ~ \cite {CH18}), in the flat one-dimensional case, there is a simple relationship between the field and the displacement
\begin{equation}
\label{peq3}
R(\rho^L,\theta) = E(\rho,\theta) \equiv E(\rho^L + R(\rho^L,\theta),\theta),
\end{equation}
which makes the functions of the electric field and displacement on the particle trajectory indistinguishable.
Note that this approach additionally makes it possible not to solve simultaneously two identical differential equations with different initial conditions (to determine the Euler trajectory of a particle and the electric field along it).
In this case, the formal transition from the Euler coordinates $ (\rho, \theta) $ to the Lagrangian coordinates $ (\rho ^ L, \tau) $ is carried out by the usual transformation
$$
\tau \equiv \theta, \quad \rho^L = \rho - \int\limits_{\tau_0}^{\tau} d\tau'V_1(\rho^L,\tau')\,,
$$
but instead of $ \tau $, the old notation $ \theta $ is used.

\subsection
{  Numerical algorithm}\label{Sec4.1}

To find a numerical solution on the line $ \rho \in (- \infty, \infty) $, we define an arbitrary grid at the initial moment of time $\theta=0$:
$$
          \rho_{1}(0) < \rho_{2}(0) < \dots < \rho_M(0),
$$
consisting of $ M $ nodes. To each node $ \rho_k (\zt = 0), \; 1 \le k \le M, $ we place the particle labeled with the Lagrangian coordinate $\rho^L_k$.
Moreover, for each particle from \eqref {3gl5non} and \eqref {peq3} we get equations describing the dynamics of particles in Lagrangian variables:
\begin{equation}
\label{peq4}
\begin{array}{c}
\dfrac{d \,V_1(\rho^L_k,\theta) }{d \,\theta} = - R(\rho^L_k,\theta) - B_0 V_2(\rho^L_k,\theta), \quad
\dfrac{d \,V_2(\rho^L_k,\theta)}{d \,\theta} = B_0 V_1(\rho^L_k,\theta), \\
\dfrac{d \,R(\rho^L_k,\theta)}{d \,\theta} = V_1(\rho^L_k,\theta),
\quad k =1, 2, \dots, M.
\end{array}
\end{equation}

Then we use  \eqref {ptraek} and \eqref {peq3} to obtain the missing initial conditions for systems  \eqref {peq4}.
At the node $ \rho_k $ for $ \zt = 0 $ the value $ E ^ 0 (\rho_k) $ is defined, namely, $ \rho_k = \rho^L_k + E^0 (\rho_k) $. Whence for the particle
with the number $ k $ we obtain the initial conditions
 \begin{equation}
\label{peq5}
\begin{array}{c}
V_1(\rho^L_k,\theta=0)= V_1^0(\rho_k), \quad V_2(\rho^L_k,\theta=0)= V_2^0(\rho_k), \vspace{1.0 ex}\\
R(\rho^L_k,\theta=0)= E^0(\rho_k), \quad k = 1,2, \dots, M,
\end{array}
\end{equation}
and the equilibrium value of the Lagrangian coordinate $ \rho^L_k $.
The obtained relations allow to solve numerically the problem \eqref {peq4}, \eqref {peq5} formulated in Lagrangian variables
instead of the problem \eqref {3gl5non}, \eqref {cd2}, written in Euler variables.

The solution of  \eqref {peq4}, \eqref {peq5} determined in this way does not allow finding the spatial derivatives of the sought functions $ V_1, V_2 $ and $ E $, which makes it impossible to determine the electron density function $ N $ in accordance with \eqref{3gl4den}. To avoid this drawback, it is sufficient to write the equations \eqref {char2d} also in Lagrangian variables, using the reasoning outlined above. Formal transformations give the equations
\begin{equation}
\label{peq6}
\begin{array}{l}
\dfrac{d \,q_1(\rho^L_k,\theta) }{d \,\theta} = - e(\rho^L_k,\theta) - q_1^2(\rho^L_k,\theta) - B_0 q_2(\rho^L_k,\theta), \vspace{1.0 ex} \\
\dfrac{d \,q_2(\rho^L_k,\theta) }{d \,\theta} = - q_1(\rho^L_k,\theta)\, q_2(\rho^L_k,\theta) + B_0 q_1(\rho^L_k,\theta), \vspace{1.0 ex} \\
\dfrac{d \,e(\rho^L_k,\theta)}{d \,\theta} = (1 -  e(\rho^L_k,\theta))\, q_1(\rho^L_k,\theta),
\end{array}
\end{equation}
and initial conditions
 \begin{equation}
\label{peq7}
\begin{array}{c}
q_1(\rho^L_k,\theta=0)= \left(V_1^0\right)'(\rho_k), \quad
q_2(\rho^L_k,\theta=0)= \left(V_2^0\right)'(\rho_k), \vspace{1.0 ex}\\
e(\rho^L_k,\theta=0)= \left(E^0\right)'(\rho_k),
\end{array}
\end{equation}
corresponding to individual particles with numbers $k = 1,2, \dots, M.$

Now we solve  problem \eqref {peq6}, \eqref {peq7} (taking into account \eqref {ptraek})
 \begin{equation}
\label{peq8}
\rho_k(\zt) = \rho^L_k + R(\rho^L_k,\theta), \quad k =1, 2, \dots, M,
\end{equation}
for each particle and find  the value of the electron density
$$
N(\rho_k, \theta) = 1 - e(\rho^L_k,\theta).
$$
at the Euler point of space $ (\rho_k (\zt), \theta) $.

Thus, the proposed numerical algorithm consists in solving equations  \eqref {peq4} with  conditions \eqref {peq5} at the nodes of the variable Euler mesh \eqref {peq8}, together with solving equations \eqref {peq6} with the conditions \eqref {peq7} for the particles $k =1, 2, \dots, M.$
Note that the construction of the approximate solution presupposes the existence, uniqueness, and smoothness of the exact solution to the problem due to Theorem \ref{T1} and general theorems on the existence and uniqueness of a locally smooth solution to the Cauchy problem for systems of hyperbolic equations  \cite {Daf16}.
If the solution has sufficient smoothness in the variable $ \zt $, it seems very convenient to use the classical Runge - Kutta method of the fourth order of accuracy  \cite {KMN89}, or schemes of a lower order of accuracy up to the Euler method if the smoothness is insufficient.
Let us stress that  the accuracy of the obtained approximation is determined exclusively by the smoothness of the solution.

Note also that the stability of the time integration of  equations \eqref {peq4} 
and \eqref {peq6}, is completely determined by condition \eqref {crit2}. In addition, this condition ensures that there are no intersections of Lagrangian trajectories, i.e. preservation of the starting order of particles, or in other words, the fulfillment of the inequality
$$
\rho_{k+1}(\zt) - \rho_{k}(\zt) > 0 \quad \forall k =1,2, \dots, M-1.
$$

Summarizing the above description of the numerical algorithm, we emphasize that to solve the original problem  \eqref {3gl5non} - \eqref {3gl4den} written in Euler variables, it is convenient to go over to Lagrangian variables and use approximate methods of time integration for the extended system of equations to find solution at the points $ \rho_k (\zt^ n), \quad k = 1, 2, \dots, M, $ $ \zt ^n = \zt^{n-1} + \tau_n, \; n = 1, 2, \dots, $ belonging to  trajectories of particles. This is quite enough for the study of most problems. However, it is quite possible that situations arise when it is required to determine the solution at the given Euler points $ (\rho, \theta) $, which do not have to belong to the calculated trajectories of the particles.

In this case, at the moment of time $ \zt^n $, it is first necessary to determine the interval $ [\rho_k (\zt^n), \rho_{k + 1} (\zt^n)] $, to which the given value of $ \rho $ belongs and then use the interpolation procedure to find the approximate value. Taking into account that at the nodes of the Euler grid $ \rho_k (\zt ^ n), \quad k = 1, 2, \dots, M, $ there are not only the values of the functions
 $ V_1 (\rho, \zt), $ $ V_2 (\rho, \zt) $ and $ E (\rho, \zt) $, but also the values of their spatial derivatives
$ q_1 (\rho, \zt), $ $ q_2 (\rho, \zt) $ and $ e (\rho, \ zt) $, it seems very convenient to use Hermitian cubic interpolation for this purpose. The derivation of the necessary formulas and error estimates is given in  \cite {Schultz73}, the practical details of use, including the necessary programs, are well described in  \cite {KMN89}. An example of a formal substantiation of such an algorithm by the example of the problem of the influence of electron - ion collisions on Langmuir oscillations of a cold plasma is given in  \cite {ChDR21}.

\vspace{1.5em}

\subsection{Examples}

We use the most natural from the point of view of physics initial condition for the electric field
\begin{equation}\label{gauss}
E^0(\rho) = \left(\dfrac{a_*}{\rho_*}\right)^2\rho \exp\left\{-2
\dfrac{\rho^2}{\rho_*^2}\right\},
\quad
\rho \in {\mathbb R},
\end{equation}
 they simulate the effect of a short high-power laser pulse when it is focused in a line (this can be achieved using a cylindrical lens), see details in \cite {Shep13}. Here $ a _ * $ characterizes the pulse power, and $ \zr_* $ is its spatial size. The value of $ \zr_* $ in the calculations was assumed to be equal to one.
As follows from \eqref {3gl4den} and \eqref {gauss}, initially the density minimum is at the origin,
|and the deviation of the density from the equilibrium state  equal to 1  decays exponentially as $|\zr|\ \to \infty$.

As the working area we select the segment $ [- d, d] $, $ d = 4.5 \, \rho_* $.
In this case, the amplitudes of particle oscillations at the boundaries will coincide in order of magnitude with the machine accuracy, since
$ \exp \{- 2d^2 / \rho_*^2 \} \approx 2.58 \cdot 10^{- 18} $, i.e. the near-boundary characteristics  coincide with the straight lines $ \rho \approx \pm d $ with a high degree of accuracy.
It is convenient to choose a uniform initial grid  $ \rho_k (\zt = 0) = kh - d, \; k = 0,1, \dots, M, \, h = 2d / M. $
In this case, the characteristic value of the spatial variable discretization parameter used in the calculations is $ h = 10^{- 3} $. For reasons of stability, the step of integration over time was chosen equal to $ h $, and in order to control the accuracy, calculations were performed regularly with grid parameters two times smaller than the main (working) ones.

A summary of the blow up effect in the framework of the non-relativistic hydrodynamic model of cold plasma is as follows. Let us fix arbitrary initial data. The initial spatial distribution of the electron density $ N $, as a consequence of the formulas \eqref {3gl4den} and \eqref {gauss}, leads to an excess of positive charge at the origin (at $ \zr = 0 $). For this reason, the electrons begin to move in the direction of the center of the region, which, after half the oscillation period, generates a density distribution with a global maximum also at $ \zr = 0 $. If the nonlinear plasma oscillations are regular (i.e. condition \eqref {crit2} holds), then the   distributions of electron density  change each other every half of the period, generating in the center of the region a strictly periodic sequence of extrema with constant amplitudes. However, if the condition \eqref {crit2} is not met, then during the first half-period, a gradual formation of a gradient catastrophe occurs for the initially smooth functions of the velocity and electric field. This can be prevented by increasing the external field in accordance with the inequality \eqref {crit2}. In other words, we can  prevent a blowup by increasing the  parameter $ B_0 $ (to stabilize the solution). However, stabilization is not possible for any choice of the initial condition $ V_2 ^ 0 (\zr) $. For example, if the choice of the initial condition ensures that the solution belongs to the subclass $ K_1 = 0 $ (see Sec.\ref{Sec3.1}), then the situation will be reversed, namely: an increase in the external field $ B_0 $ will lead to an acceleration of the blow up process in time.
Nevertheless, the parameter $B_0$ participate in the initial condition $ V_2 ^ 0 (\zr) $ and increasing $B_0$ we change the initial data. The situation is different from the one described above when we study the increase in $ B_0 $ with fixed initial data.

Let us proceed to the  numerical experiments.

\begin{center}
\begin{figure}[htb]
\begin{minipage}{0.495\columnwidth}
\includegraphics[scale=0.6]{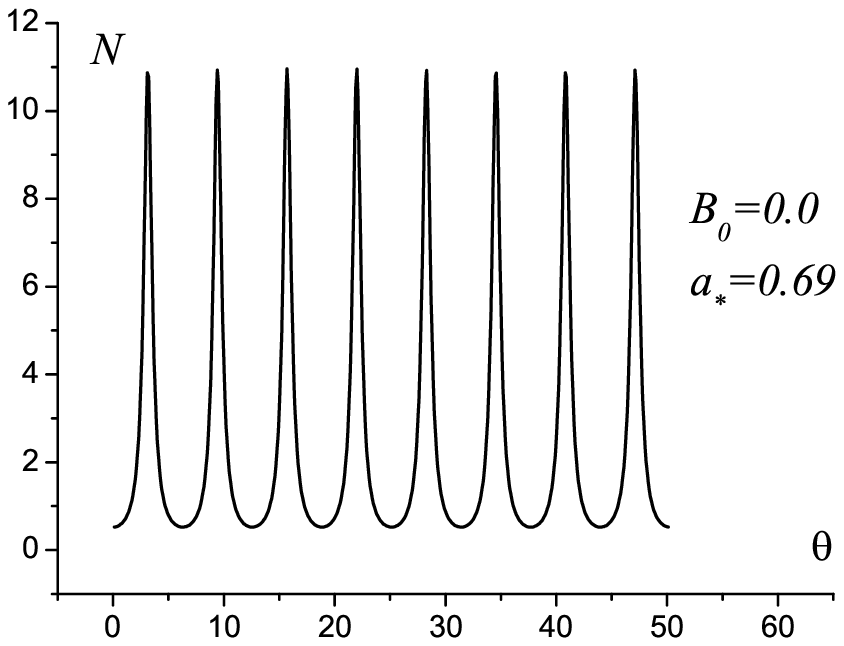}
\end{minipage}
\begin{minipage}{0.495\columnwidth}
\includegraphics[scale=0.6]{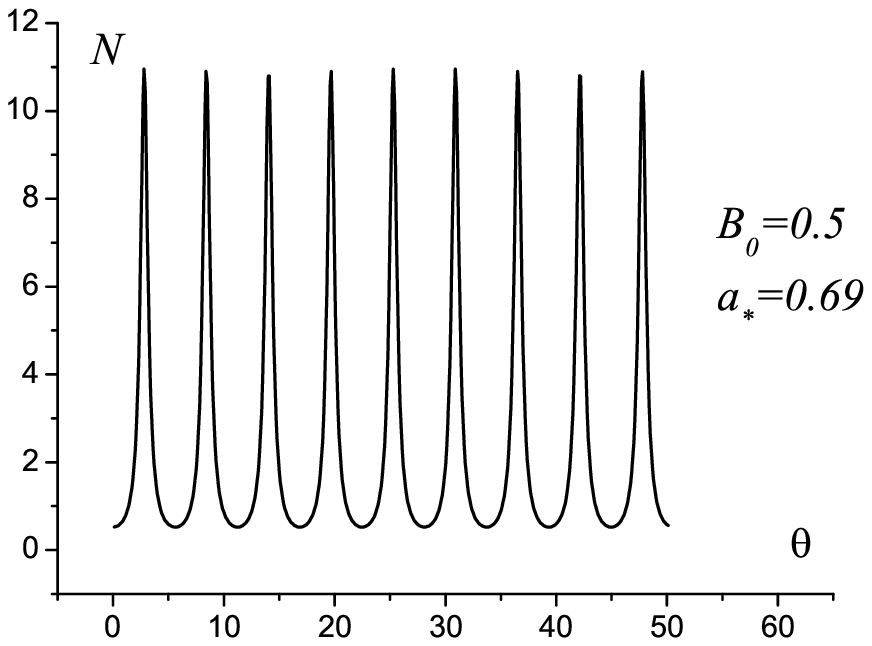}
\end{minipage}
\caption{The behavior of electron density $ N (\rho = 0, \theta) $ in the absence (left) and in the presence (right) of an external magnetic field (regular oscillations).}\label{Pic2}
\end{figure}
\end{center}

1. First, we present some illustrations of smooth solutions in the case where we have a possibility to compare the numerical result with an exact solution. We choose $ V_1^0 (\zr)=0 $ and note that the exact value of the density at the origin can be found for the class of solutions $K_1=0$ from \eqref{char2d}, \eqref{C1} for this case, where the characteristic line $\zr=0 $ does not depend on time. Namely,
$$
N(\rho=0,\theta) = \dfrac{1-\alpha}{1-\alpha(1-\cos  \sqrt {1 + B_0 ^ 2}\theta)}, \quad \mbox{\rm with} \quad
\alpha = \left( \dfrac{a_*}{\rho_*}\right)^2, \; \zr_*=1.
$$
For the case $ B_0 = 0 $ and initial conditions \eqref {gauss} with $ a_* = 0.69 $, condition \eqref{crit2} holds and the solution  exists for all $t>0$. The electron density here is $ 2 \pi- $ periodic in time and it oscillates in the approximate range of $ [0.52,10.96] $.
Fig.\ref{Pic2}, left, shows the graph of $ N (\rho = 0, \theta) $, obtained using the numerical algorithm described in Sec.\label{Sec4.1}. It is not possible to distinguish between a visually approximate solution and an exact one.

 Then we put $ B_0 = 0.5 $ and simultaneously change the initial condition to
\begin{equation}
\label{sclK0}
V_{2}^0(\rho) = B_0 E^0(\rho), \quad \rho \in {\mathbb R},
\end{equation}
to fall into the class $K_1=0$. The condition \eqref {crit2} is satisfied; therefore, the numerical algorithm generates a regular solution, the electron density from which is shown in Fig.\ref{Pic2}, right. It coincides with the exact one.
It is easy to see that the oscillation frequency (equal to $ \sqrt {1 + B_0 ^ 2} $) has increased, therefore, at the same time interval, we observe 9 full oscillation periods instead of 8 in the previous  case $ B_0 = 0 $.

\begin{center}
\begin{figure}[htb]
\begin{minipage}{0.495\columnwidth}
\includegraphics[scale=0.6]{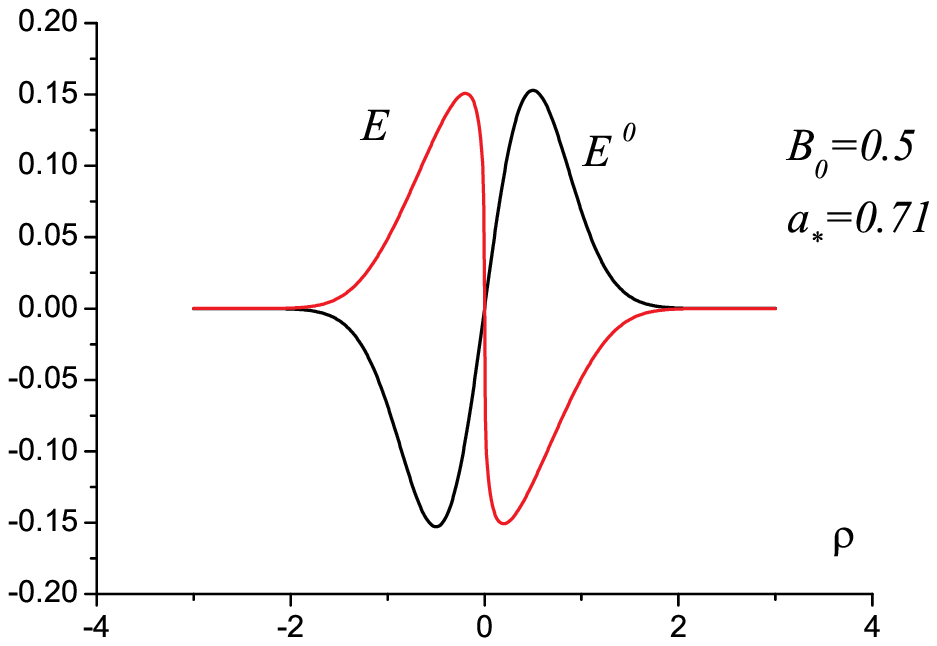}
\vspace{-0.5 cm}
\end{minipage}
\begin{minipage}{0.495\columnwidth}
\includegraphics[scale=0.6]{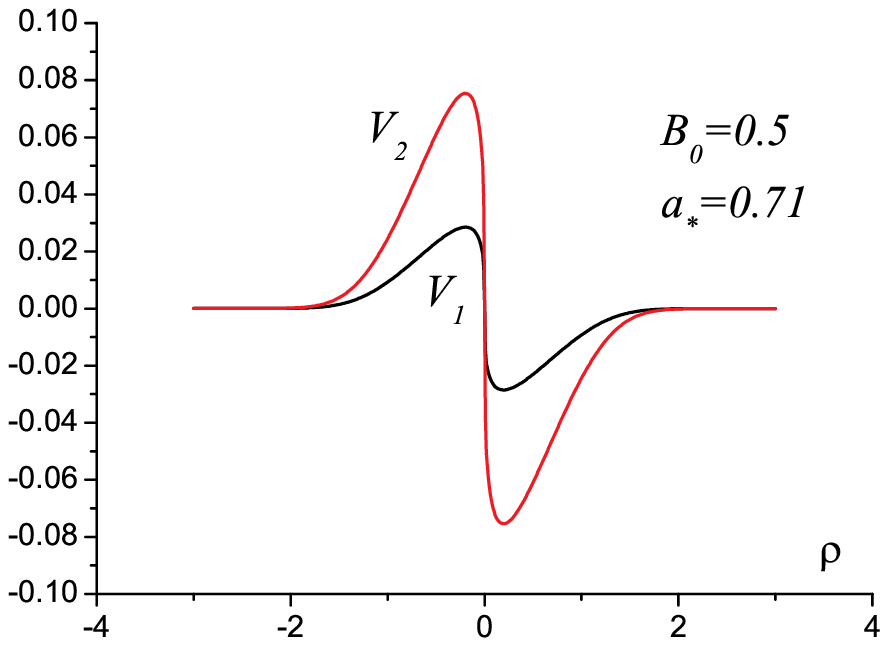}
\vspace{-0.5 cm}
\end{minipage}
\caption{Formation of the gradient catastrophe. Transformation of the smooth initial condition $ E^0 (\zr) $ into the step function $ E (\zr) $ (left) and the blow up of velocity components $ V_1 $ and $ V_2 $ (right).}\label{Pic3}
\end{figure}
\end{center}

2.
Let us choose $ a_* = 0.71 $ with the same parameters $ \zr_* = 1, \, B_0 = 0.5 $ and the initial condition \eqref {sclK0}).
Now the condition \eqref {crit2} is violated and numerical calculations  lead to the blow up effect described above. The gradient catastrophe is shown in Fig.\ref{Pic3}. Fig.\ref{Pic3}, left, shows the transformation of the smooth initial condition $ E ^ 0 (\zr) $ into the step function $ E (\zr) $, this happens at the moment of time $ \zt \approx 2.66 $. It is not possible to depict the electron density function $ N $ here, due to its singularity at the point $ \zr = 0 $.
The components $ V_1 $ and $ V_2 $ are shown in Fig.\ref{Pic3}, right. They also turned into step functions from smooth initial functions defined by the relations   \eqref{gauss}  and \eqref{sclK0}.
Let us note that for $ \zr_* = 1, \, B_0 = 0 $ and $ a_* = 0.71 $, the  condition \eqref {crit2} is also violated. With these parameters, the breaking effect is observed earlier in time ($ \zt \approx 2.98 $). And this is a clear illustration of the fact that if the solution belongs to the subclass $ K_1 = 0 $, then an increase in the external magnetic field not only does not "smooths" the blow up effect, but, on the contrary, brings it closer in time.

3. Let us further demonstrate the process of stabilization the solution by increasing the external magnetic field. To demonstrate a stabilization, we replace  the initial condition \eqref {sclK0} with
\begin{equation}
\label{init0}
V_{2}^0(\rho) = 0, \quad \rho \in {\mathbb R},
\end{equation}
which does not fall into the class $K_1=0$.

\begin{center}
\begin{figure}[htb]
\hspace{-1.5cm}
\begin{minipage}{0.4\columnwidth}
\includegraphics[scale=0.6]{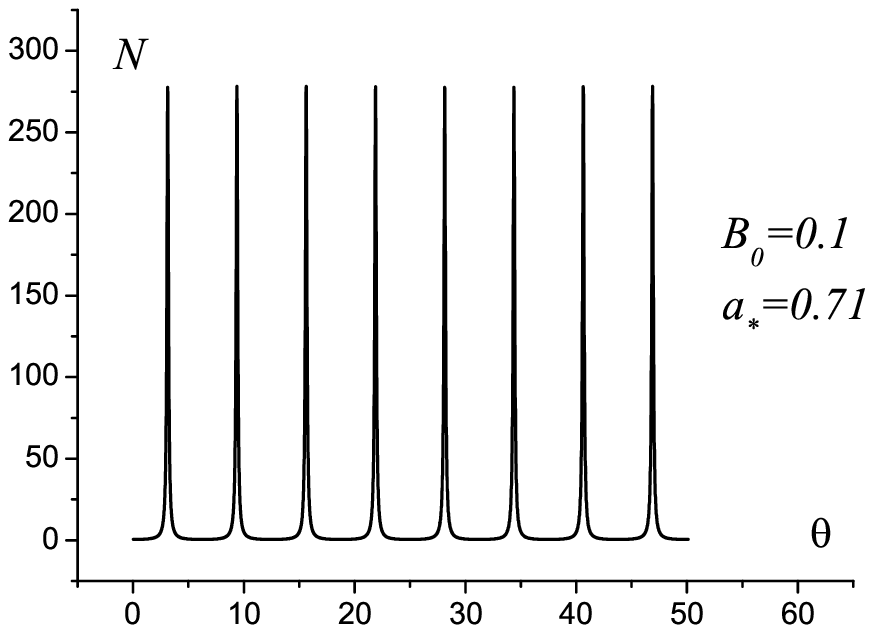}
\end{minipage}
\hspace{1.5cm}
\begin{minipage}{0.4\columnwidth}
\includegraphics[scale=0.6]{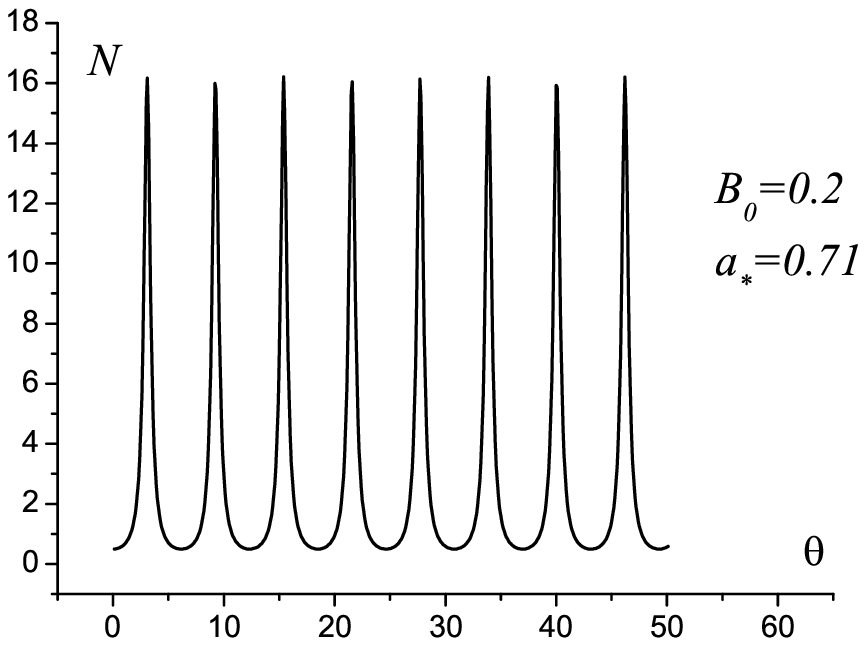}
\end{minipage}
\begin{minipage}{0.4\columnwidth}
\includegraphics[scale=0.6]{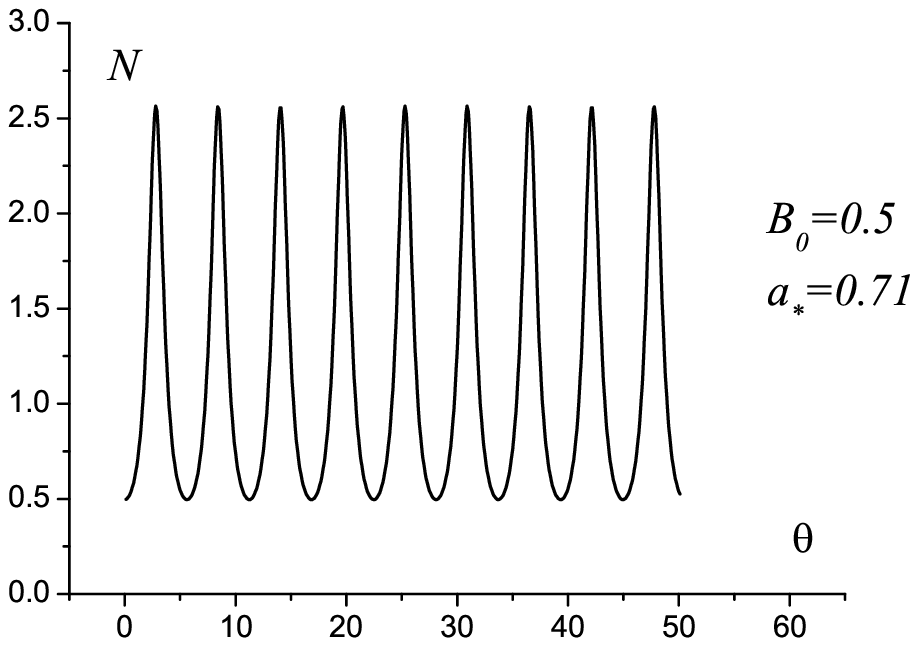}
\end{minipage}
\caption{Stabilization of the solution. The electron density $ N (\rho = 0, \theta) $ at a weak (left), moderate (right) and strong (below) external magnetic field.}\label{Pic4}
\end{figure}
\end{center}

The result of  calculation for the parameters $ \zr_* = 1, \, B_0 = 0.1 $, $ a_* = 0.71 $ and the initial condition \eqref {init0}
is shown in Fig.\ref{Pic4}, left. The density is uniformly bounded, although the extreme values are large enough ($ \approx $ 275).

It should be borne in mind that for the values of the magnetic field $ 0 \le B_0 \le B_ {crit} <0.1 $ (with the remaining parameters fixed), the condition \eqref{crit2} is violated and the  solution blows up.

An increase in the external field to $ B_0 = 0.2 $ (while maintaining the remaining parameters) leads to a decrease in the amplitude of oscillations.
 However,  the external field has not yet made a noticeable change in the oscillation frequency; therefore, in Fig.\ref{Pic4}, right, we observe only 8 complete periods.



Finally, for $ B_0 = 0.5 $, 
there is clearly a strict periodicity in time and a noticeable change in frequency (9 periods, as follows from Theorem \ref{T1}) and a strong decrease in the amplitude of oscillations (by about a factor of 6), Fig.\ref{Pic4}, below.

Note that the above results of calculations by the particle method were completely reproduced by additional (control) calculations according to the McCormack-type scheme of the second order of accuracy in Euler variables (see details in \cite {FCh21MM}).

\vspace{1.5em}
\section
{Conclusion}

In the present work, the effect of an external magnetic field on plane nonrelativistic nonlinear plasma electrostatic oscillations generated by a powerful laser pulse is investigated analytically and numerically. A necessary and sufficient condition for the existence of a smooth solution of the corresponding hyperbolic system of equations is obtained.
It is shown that with an increase in the intensity of the external magnetic field, the set of initial data corresponding to a globally smooth solution  extends. To analyze the regularizing effect of an external field, we analyzed the subclasses of solutions to the problem on which the reduction of the number of equations is allowed, including  traveling waves.
For the purpose of numerical simulation, an algorithm is constructed in Lagrangian variables, which is closely related to the method of characteristics for the original differential setting.

Numerical experiments clearly illustrate both the dynamics of smooth solutions and the formation of a gradient catastrophe. Additionally, calculations of the stabilizing effect of an external magnetic field are presented. The results obtained can be used to simulate upper hybrid oscillations at relativistic electron velocities, i.e. to study the influence of an external magnetic field on the breaking effect of multi-period plasma oscillations. In addition, it is possible to extend the results to the case of taking into account electron - ion collisions.

\section*{Acknowledgment}
Supported by the Ministry of Education and Science of the Russian Federation as part of the program of the Moscow Center for Fundamental and Applied Mathematics under the agreement ¹075-15-2019-1621.

\end{document}